\documentclass[runningheads]{llncs}
\usepackage{graphicx}
\usepackage{array, booktabs}
\usepackage{multirow}

\usepackage{makecell}

\begin{document}
\title{A Hybrid Execution Environment for Computer-Interpretable Guidelines in PROforma}
\titlerunning{A Hybrid Execution Environment for PROforma CIGs}
%
\author{Alexandra Kogan\inst{1}\orcidID{0000-0002-5926-1606} \and
Roy Leizer\inst{1}\orcidID{0009-0000-3149-4427} \and
Szymon Wilk\inst{2}\orcidID{0000-0002-7807-454X} \and
David Glasspool\inst{3}\orcidID{0000-0001-7688-8251}}
\authorrunning{A. Kogan et al.}
%
\institute{University of Haifa, Haifa, Israel\\
\email{akogan09@campus.haifa.ac.il} \\
\and
Poznan University of Technology, Poznan, Poland\\
\and
Deontics Ltd., London, United Kingdom\\
}
\maketitle              
\begin{abstract}
In this paper, we share our experience of developing a hybrid execution environment for computer-interpretable guidelines (CIGs) in PROforma. The proposed environment is part of the CAPABLE system which provides coaching for cancer patients and decision support for physicians. It extends a standard PROforma execution engine -- Deontics Engine (DE) -- with additional components that act as wrappers around DE, allow handling non-standard tasks and facilitate integration with the rest of the CAPABLE system. This yields a hybrid environment in which the standard engine and specialized components must be interfaced together by some intervening layer. In the CAPABLE system this has been achieved by defining a set of specialized meta-properties which are attached to data and tasks in the PROforma CIGs to specify the interface between engine and components.

\end{abstract}
\section{Introduction}

The CAPABLE (CAncer PAtients Better Life Experience) system aims to provide \textit{coaching} for cancer patients treated at home and \textit{clinical decision support} for their physicians \cite{capable-demo-2021}. Physicians receive recommendations about cancer treatment-related symptoms, as well as alerts and notifications about circumstances requiring their attention. Patients are provided with coaching that covers monitoring of reported symptoms, recommending pharmacological treatments (only if pre-authorized by the physician) and non-pharmacological (so-called \textit{virtual capsules} addressing psychological and physical well-being \cite{capsules-ideas-2021}) interventions. The system also produces alerts, reminders, and various types of tips (e.g., related to prevention, symptom management or motivational support). 

When providing support to patients and physicians, CAPABLE relies on data- and knowledge-driven models, and in this paper we focus on the latter. These models represent clinical practice guidelines (CPGs), workflows (both administrative and clinical), and simpler rules. For simplicity, in the following text, we refer to these models as computer-interpretable guidelines (CIGs), although some of them are not directly concerned with evidence-based management of specific conditions. We represent CIGs using the PROforma formalism \cite{sutton-2003} that is based on a task-network model, where possible tasks include enquiries, decisions, and actions. We decided to use PROforma due to its expressiveness, availability of authoring and execution tools, and experience in earlier projects \cite{kogan-2020}.  

The Deontics Engine (DE) -- an execution engine for PROforma -- provides a stand-alone execution environment with an internal database and a web-based user interface (UI) for entering data, presenting recommendations on actions and decisions (with optional explanations), and collecting user feedback on the provided recommendations. However, it also exposes a REST API that allows external components to interact with the engine, retrieve and set values of data items and control the status of the engine, processed CIGs, and individual tasks. 

In CAPABLE we took advantage of that API and developed a \textit{hybrid execution environment} for PROforma CIGs that extends DE with the following components: Physician Decision Support System (PDSS), Virtual Coach (VC), and a Goal-Oriented Comorbidities Controller (GoCom) \cite{kogan-2020}. PDSS and VC act as wrappers around DE that are able to handle non-standard tasks and mediate access to a shared, FHIR-based \cite{hl7-fhir} data repository. GoCom is responsible for the detection of interactions and providing explanations and justifications for alternative treatments. Finally, a component that is adjacent to the hybrid CIG-execution environment, the Knowledge-Data Ontology Mapper (KDOM) provides logical and temporal abstractions derived from data, thus simplifying the definition and execution of CIGs. 

The DE provides a standardised view of the CIGs. However this needs to be interfaced to the specific architecture and data representations of the patient management system it is embedded into. PROforma allows meta-properties to be associated with all tasks and data for this purpose. For CAPABLE 
an extensive set of meta-properties was developed, which we believe may prove useful in other systems.
In the following sections, we present this hybrid environment - standardised CIG engine, specialized components and intervening meta-properties - with a focus on the PDSS, VC and GoCom components.

\section{High-level Architecture of the Hybrid Environment}

\begin{figure}[t]
\centering
\includegraphics[scale=0.35]{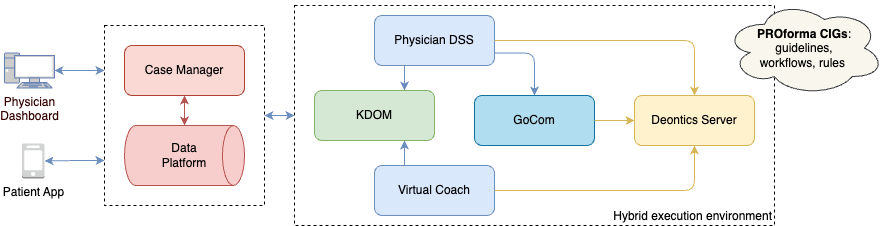}
\caption{A high-level architecture of the CAPABLE system} 
\label{fig:capable-architecture}
\end{figure}

In Fig. \ref{fig:capable-architecture} we present a high-level architecture of CAPABLE and mark the components that comprise the hybrid execution environment. The Data Platform (DP) and Case Manager (CM) provide a FHIR-based data and event management platforms (for more details see \cite{polce-2021}). Here, we should note that CM acts similarly to the Clinical Events Monitor introduced in the GLEE system as part of the electronic health record interface \cite{glee-2004}. Moreover, the Patient App and Physician Dashboard implement a mobile UI for patients and a web-based UI for physicians, respectively. 

Blue arrows in Fig. \ref{fig:capable-architecture} indicate the conceptual exchange of information and control messages between components and are implemented using more complex protocols. Interactions between PDSS, GoCom, VC, and DE, marked with orange solid arrows, are handled differently and employ the CIG-related REST API. DP and CM also mediate interactions between the proposed CIG execution environment and the UI components. Recommendations generated by PDSS and VC for physicians and patients, respectively, are stored in DP (as \textsf{Communication} resources), and appropriate UI components are notified by CM.


\section{Execution of PROforma CIGs}

PDSS and VC rely on the core functionality provided by DE, e.g., traversal of CIGs according to scheduling constraints or evaluation of decision candidates, and  adopt dynamic instantiation and execution of CIGs. Specifically, CIG instances are created in response to selected events reported by CM (e.g., a new symptom entered by a patient) and terminated as soon as these events have been handled. Relevant data related to CIG execution (e.g., logs of completed tasks and their results) is stored in DP -- PDSS and VC do not need to keep it in memory thus increasing their robustness, as component restarts do not require recovering or rebuilding their internal state. Such a solution is similar to execution traces employed by the GLEE system \cite{glee-2004}.

\begin{figure}
\centering
\includegraphics[scale=0.35]{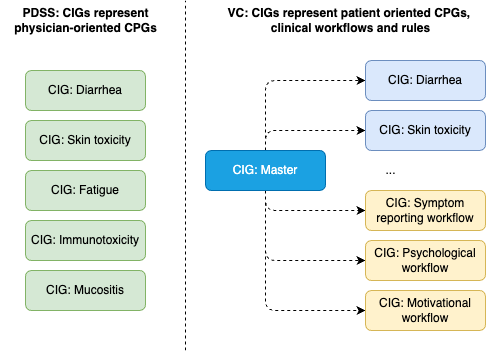}
\caption{CIGs used in CAPABLE and their organization}
\label{fig:cigs}
\end{figure}

There are differences in the organization of CIGs employed by PDSS and VC, presented in Fig.~\ref{fig:cigs} and in operational cycles adopted by both components. PDSS uses 5 CIGs that represent patient-oriented parts of ESMO (European Society of Medical Oncology) CPGs for managing specific toxicities caused by cancer treatment. All these CIGs are complex and share a common structure covering data collection, diagnosis, prevention, treatment and monitoring stages. Moreover, they all include applicability conditions that are checked after the data collection stage to ensure that only relevant actions are activated and performed. 

On the contrary, CIGs used by VC represent not only patient-oriented parts of ESMO CPGs, but also workflows and rules corresponding to various forms of support, e.g., psychological and motivational, or for managing symptom reporting. Their number is much larger (currently 40), they do not have a fixed structure and they can be invoked in many different contexts. To streamline the definition and maintenance of their applicability conditions, we introduced the master CIG that is invoked first. It checks the current context (e.g., patient clinical status, current time), identifies applicable (possibly multiple) specific CIGs, and invokes these CIGs.

\subsection{An Operational Cycle of PDSS}

\begin{figure}[t]
\centering
\includegraphics[scale=0.32]{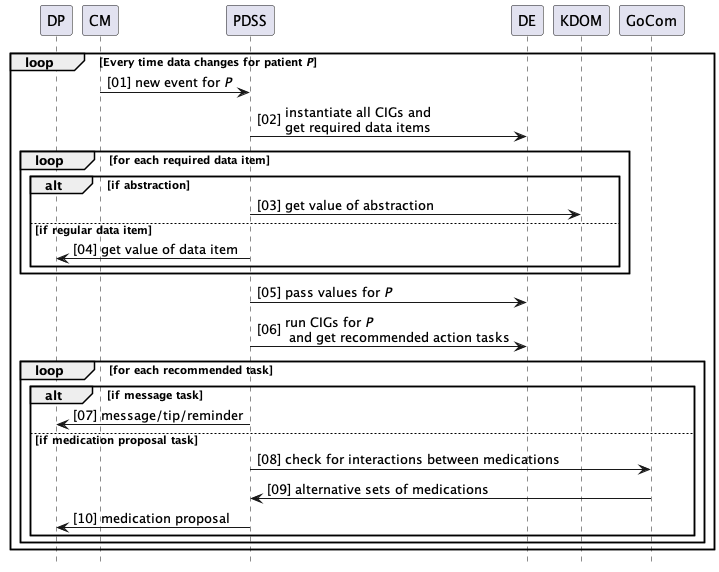}
\caption{An operational cycle of PDSS (specific interactions between components are described in the text)}
\label{fig:pdss-functional-cycle}
\end{figure}

In Fig. \ref{fig:pdss-functional-cycle} we present a UML sequence diagram illustrating an operational cycle adopted by PDSS that relies on repeated \textit{assessments} of a patient. An assessment is initiated by an event reported by CM (interaction 01 in Fig. \ref{fig:pdss-functional-cycle}, later in the text we refer to other interactions in this figure). PDSS instantiates all available CIGs for a given patient and retrieves data items that are required by these CIGs (interaction 02). Then, it iterates over data items and either asks KDOM to compute abstraction values or retrieves values from DP (interactions 03--04). The source is specified by one of the specialized meta-properties (see Section \ref{sec:meta-props}). The collected values are passed to DE which runs the instantiated CIGs until their completion and reports to PDSS the list of recommended action tasks (interactions 05--06). Since these CIGs internally check for applicability conditions, only relevant tasks are reported to PDSS.

In the second cycle phase, PDSS iterates over recommended tasks and depending on their type and meta-properties, either passes a simple recommendation (e.g., a tip or reminder) to the Physician Dashboard via DP (interaction 07), or in the case of medication proposals it interacts with GoCom to mitigate potential conflicts (interactions 08--09) and passes the revised proposal to the Dashboard (interaction 10). Specifically, GoCom receives notifications from PDSS regarding recommended medications. It checks each medication in the RxNav database for interactions with the patient's active medications and recommendations from other guidelines. GoCom attaches textual evidence from the guideline and if an interaction is found, generates an additional explanation. Finally, the recommendations are formatted based on the logical gate of the decision that indicates which recommendations should be followed (see Fig. \ref{fig:task-props}), and the final response is sent back to PDSS.

\subsection{An Operational Cycle of VC}

\begin{figure}
\centering
\includegraphics[scale=0.32]{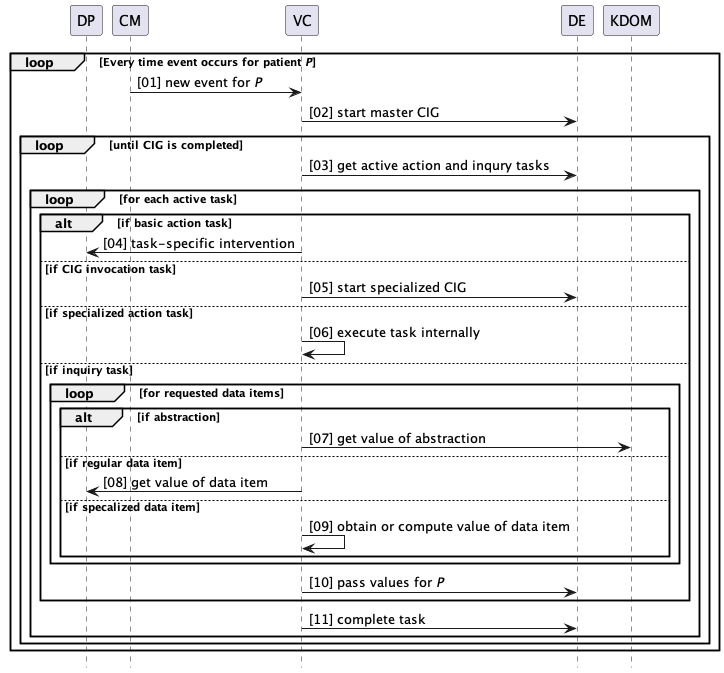}
\caption{An operational cycle of VC (specific interactions between components are described in the text)}
\label{fig:vc-functional-cycle}
\end{figure}

In Fig. \ref{fig:vc-functional-cycle} we demonstrate a UML sequence diagram with the operational cycle of VC. It is more complex than the one for PDSS, as VC needs to control the execution of individual tasks. When CM reports an event (interaction 01 in Fig. \ref{fig:vc-functional-cycle}, later in the text we refer to other interactions from this figure), VC instantiates and starts the master CIG for a given patient (interaction 02). Then, it enters the loop when it checks for active action and inquiry tasks (interaction 03) and iterates over these tasks.

In addition to a basic action (interaction 04), VC may also invoke a specialized CIG (interaction 05) and execute an internal task (interaction 06) that implements more advanced processing  (e.g., scheduling multiple motivational tips). Moreover, in the case of inquiry tasks, in addition to KDOM and DP (interactions 07 and 08, respectively) VC also allows for querying other data sources (e.g., a repository with prevention tips) or running additional calculations (interaction 09). Collected values are passed to DE and the current task is completed (interactions 10--11).  

\section{Specialized PROforma Meta-properties}
\label{sec:meta-props}

Meta-properties in PROforma allow structured information to be associated with a data item or a task. Such information is not used by DE but can be accessed by external systems. We have used this feature to bridge between the DE and the PDSS, VC and GoCom components. 
In Fig. \ref{fig:data-item-props} and \ref{fig:task-props} we list some examples of the meta-properties associated with data items and tasks, respectively, that were developed for the hybrid CIG execution environment in the CAPABLE system.

\begin{figure}[!htb]
    \begin{footnotesize}
    \begin{itemize}
        \item \textsf{sourceType}
        \begin{itemize}
            \item \textit{Description}: indicates the component that created the data item. If not specified, a data item created by any component can be retrieved.
            \item \textit{Example}: \textsf{patient} (Patient App)
        \end{itemize}
         \item \textsf{resourceType}
         \begin{itemize}
             \item \textit{Description}: indicates the type of FHIR resource to be used as the source for data values. 
             \item \textit{Example}: \textsf{Communication}
         \end{itemize}
        \item \textsf{valueExpression}
        \begin{itemize}
            \item \textit{Description}: Specifies what property of a FHIR data item should be reported to DE.
            \item \textit{Example}: \textsf{cancer-treatment-related} (a flag in an \textsf{Observation} resource where the physician indicates if the symptom is related to the cancer treatment)
        \end{itemize}        
    \end{itemize}
    \end{footnotesize}
    \caption{A list of selected meta-properties characterizing PROforma data items}
    \label{fig:data-item-props}
\end{figure}

\begin{figure}[!htb]
   \begin{footnotesize}
    \begin{itemize}
        \item \textsf{interventionType}
        \begin{itemize}
            \item \textit{Description}: indicates intervention represented by a task.
            \item \textit{Example}: \textsf{medication-proposal} (a pharmacological treatment recommendation proposed by the system)
        \end{itemize}
        \item \textsf{gate}
        \begin{itemize}
            \item \textit{Description}: determines the number of recommendations to follow when multiple options are offered by the system.
            \item \textit{Example}: \textsf{AND} (all recommendations), \textsf{OR} (at least one recommendation), \textsf{XOR} (exactly one recommendation)
        \end{itemize}
    \end{itemize}
    \end{footnotesize}
    \caption{A list of selected meta-properties characterizing PROforma tasks}
    \label{fig:task-props}
\end{figure}

\section{Conclusion}

This paper has presented a novel hybrid CIG-execution environment within the CAPABLE system, focusing on the architecture of the environment and operational cycles adopted by specific components. The extension of the functionality of DE with the PDSS, VC and GoCom components as well as the introduction of an extensive set of custom meta-properties has allowed us to facilitate the mapping of patient data among multiple components and processes, optimize the interactions between the components and address challenges related to CIG execution, non-standard task handling, and conflict mitigation between proposed and prescribed pharmacological treatments facilitated by GoCom. 

The proposed environment has been implemented using programming languages for Java VM -- Scala (VC) and Java (the remaining components). All components are multi-threaded (they use either classical Java threads or an actor-based model and the Akka framework) and they are able to simultaneously handle multiple patients managed according to multiple CIGs. The environment was tested using synthetic realistic scenarios based on actual clinical cases and a dedicated simulation environment \cite{glasspool-2022}. Now, the CAPABLE system is now undergoing clinical tests in two institutions.

\subsubsection*{Acknowledgments}
This work has received funding from the EU's Horizon 2020 research and innovation programme under grant agreement No 875052.

\bibliographystyle{splncs04}
\bibliography{refs}
%




\end{document}